\documentclass[manuscript]{aastex}
\slugcomment{}
\shorttitle{Mid-infrared Observations of Three Seyfert Galaxies}
\shortauthors{Soifer et al.}

\begin{document}

\title{
High Spatial Resolution Mid-infrared Observations of Three Seyfert Galaxies 
}
\author{B.~T.~Soifer\altaffilmark{1,2},
~J.~J.~Bock\altaffilmark{3,4},~K.~Marsh\altaffilmark{5},
~G.~Neugebauer\altaffilmark{1,6},~K.~Matthews\altaffilmark{1},
~E.~Egami\altaffilmark{1,6},~L.~Armus\altaffilmark{2},
}
\altaffiltext{1}{Caltech Optical Observatories, California Institute of
Technology,
105-24, Pasadena, CA 91125}
\altaffiltext{2}{SIRTF Science Center, California Institute of
Technology
314-6, Pasadena, CA 91125}
\altaffiltext{3}{Division of Physics, Mathematics and Astronomy, California Institute of Technology,
Pasadena, CA 91125}
\altaffiltext{4}{Jet Propulsion Lab, 169-327, 4800 Oak Grove Dr., 
Pasadena, CA 91109}
\altaffiltext{5}{IPAC,Jet Propulsion Lab/ California Institute of Technology,
100-22, Pasadena, CA 91125}
\altaffiltext{6}{present address:Steward Observatory, University of Arizona,
Tucson,AZ 85721}
\email{ bts@irastro.caltech.edu,jjb@astro.caltech.edu, kam@ipac.caltech.edu,
gxn@caltech.edu, kym@caltech.edu, eegami@as.arizona.edu,lee@ipac,caltech.edu 
}

\begin{abstract}
Images at 12.5~$\mu$m of nuclei of three nearby Seyfert galaxies
--- NGC~1275, NGC~4151 and NGC~7469 --- have been obtained with
the Keck 10-m Telescope. NGC~7469 is resolved and deconvolution
delineates a structure ($<$0.$''$04~)$\times$~0.$''$08 or
$<$13~$\times$~26 pc at a position angle of 135 $\degr$. From a
comparison with structure seen at millimeter  wavelengths, this
structure is interpreted as a disk aligned with the molecular
gas in the central few hundred parsecs of the galaxy. NGC~1275
and NGC~4151 are not resolved; limits on the sizes of these
nuclei are 0\farcs08 and 0\farcs16, corresponding to physical
spatial scales of 28 and 10~pc. The lower limits to the
brightness temperatures implied by these size limits and the
measured flux densities are within $\sim$50~K of the
12~-~25~$\mu$m color temperatures of these systems as inferred
from IRAS observations.  The angular size limits are within a
factor of 2.5 of the sizes required to spatially resolve thermal
emission from dust heated by a central luminosity source. These
sizes preclude significant contributions to the nuclear infrared
emission from star forming regions. 
\end{abstract}

\newpage
\section{Introduction} 
There is much evidence (e.g., Krolik 1999) that nonthermal 
processes power the nuclei, i.e., the central sources, of
Seyfert galaxies.  In the mid-infrared, however, thermal
radiation from heated dust contributes an additional emission
component. This mid-infrared radiation, at characteristic
temperatures of a few hundred kelvin, is believed to emerge from
a disk-like structure that is heated by the inner accretion disk
(Sanders et al.\ 1989). A question remains of the geometry of
the central, presumably mainly nonthermal, source and how much
of the surrounding dust is  heated by the central source and how
much is heated by starbursts which are located throughout the
nucleus and thus form an extended cloud. It is difficult to
resolve this dust component in direct observations using
currently available telescopes because the angular extent of the
region containing the emitting  dust is small. Quantitatively,
the  Seyfert galaxies considered here have observed luminosities
of 10$^{10}$-10$^{11}$L$_{\sun}$. If the dust is at
a  temperature of $\sim$300~K --- the temperature at which
blackbody emission peaks at 12.5~$\mu$m ---and is heated by a
central source, it is predicted to have a separation from the
central source 2 - 20 pc or $<$~0$''$.1 depending on the
composition of the dust. If, on the other hand, the dust were
heated by hot stars in starburst regions, the extent of the
region containing dust clouds could be much larger.

The diffraction limited full width at half maximum (FWHM) of the
Keck 10-m Telescope --- the largest aperture telescope presently
available ---  is 0$''$.31 at 12.5~$\mu$m. Deconvolution
techniques and proper sampling of a nearby point source function
(PSF) can improve this resolution significantly. For example,
deconvolution applied to the image of NGC~1068 --- the nearest
and brightest Seyfert galaxy --- have successfully  resolved
structure with a size of 0$''$.1 ($\sim$7~pc at the distance of
NGC~1068, Bock et al.\ 2000).

In this paper we present observations of the nuclei of three
nearby, infrared  bright Seyfert galaxies --- NGC~1275, 
NGC~4151, and NGC~7469 --- taken with the Keck 10-m Telescope.
These observations, when deconvolved,  approach the angular resolution
necessary to distinguish  thermal from  nonthermal emission and,
in one case, resolve the structure of the central source.
In  the mid-infrared,  all three galaxies are known to contain
a  strong point-like nucleus and  the observations were
designed,  and optimized, to probe this source. As a result,
they are not  optimized to delineate low surface brightness,
extended emission. They are thus complementary to  observations
such as those by Radomski et al.\ (2003) of NGC~4151. We take
the Hubble constant to be 75 km s$^{-1}$Mpc$^{-1}$.

\section{Sample}
The characteristics of the three Seyfert galaxies (Seyfert 1943), taken from
the literature, are shown in Table~1. All are known to be bright
in the mid-infrared (IRAS Point Source Catalog 1989; Miles et
al.\ 1996). NGC~1275 and NGC~4151 are both
compact in the mid-infrared. Miles et al.\  have shown, from
observations at $\sim$10~$\mu$m with 0$''$.7 resolution, that
NGC~7469 consists of a compact nucleus surrounded by a lumpy
ring $\sim$3$''$ in diameter. The ring is presumably a region of
active star formation. Miles et al. found that the nucleus plus
ring structure in NGC~7469 accounts for $\sim$60\% of the total
12~$\mu$m flux density of the galaxy measured by IRAS.

\section{Observations}
 The observations were made using the Keck Long Wavelength
Spectrometer (LWS; Jones \& Puetter 1993) in the imaging mode at
12.5~$\mu$m  ($\Delta\lambda$= 1.16~$\mu$m) at the f/25 forward
Cassegrain focus of the Keck I 10-m Telescope. The LWS Si:As array has
128 $\times$ 128  pixels. The scale is
0$''$.08~pixel$^{-1}$ giving a field of
$\sim$10$''~\times$~10$''$. A log of the observations is given
in Table~2.

In the mid-infrared there were no stars in the field from which
to  estimate  point source profiles. Each measurement of the 
object size was therefore accompanied by multiple
observations of one of the three bright stars, identified as the
PSF in Table~2, which was nearby to the object and of comparable
infrared brightness.  The observations were made in runs of 27
-- 54~s by interleaving imaging of the galaxy with observations
of the PSF star; the number of comparisons between the galaxy
and the PSF star is also indicated in Table~2. The telescope was
guided during the observations of the objects using a visual
guider on stars near the  galaxies. 

For these observations, objects and PSF calibrators were imaged
in a  chop-nod scheme where the throw of the secondary chopper 
was equal to the amplitude of the (telescope) nod. The amplitude 
of the secondary chopper was set to  5$''$ -- 15$''$ at a frequency 
$\sim$5~Hz.   

All three nights were photometric and in all cases the airmass
was between 1.0 and 1.4. Each sequence to measure the object
size  was preceded by an observation of one of the three bright
stars ~---~$\alpha$ Tau=HR~1457, $\alpha$ Boo=HR~5340, or 
$\beta$ Peg=HR~8775~---~ in order to obtain a photometric
calibration. 

\section{Data Reduction and Analysis} 
\subsection{Photometry}
Photometric measurements were made from the raw images before
any deconvolution algorithms (described below) were applied. The
photometry of the galaxies was based on the following magnitudes
(in the Vega based system)
--- HR~1457: [12.5~$\mu$m]=-3.07~mag,
HR~5340:~$[12.5~\mu$m]=-3.15~mag, and 
HR~8775:~$[12.5~\mu$m]=-2.55~mag. The ``sky'' was taken as an
annulus 4\farcs6 to 6\farcs0 in diameter. The conversion from
magnitude to flux density followed the prescription given in the
Explanatory Supplement to the IRAS Catalog (Beichman et al.\
1985); i.e., at 12.5~$\mu$m, 0.0~mag  was taken as 26.15~Jy. 

\subsection{General Image Data Reduction} 
The images were processed following a similar procedure to that
described in  Bock et al. (2000), briefly summarized as follows.
Images were obtained by shifting and combining sub-images after
every chop pair, using only the central image, to eliminate any
possibility of non-reproducibility in the telescope nod, i.e.,
potential differences caused  by guiding the telescope were
obviated in software by finding the centroid of each object 
image in each nod. Both end chop images did not always fall
onto the array, so were not used in the data reduction.  The
results when  the end images were included, however, were found 
to be appreciably the same. 

For the PSFs, which were not guided, each chop-nod  combination
consisted of four separate  images, which were combined to
produce three images: an image near  the middle of the  array
and two images near the edge or off the array.  The middle 
image was thus seen twice, first in the positive beam of the
first chop  pair, and then in the negative beam of the second
chop pair. The objects were guided, and eight sub-images were
combined into the three images; the middle  image was thus
seen four times.The object and  PSF calibrator images were
sub-pixel shifted, rotated, and co-added  using a routine  that
relies on sub-Nyquist  sampling of the images.  The objects were
rotated from array  coordinates to equatorial north coordinates
during the combination.   The PSF calibrators were rotated
through a sequence of rotations in  array coordinates matched to
the object.  The rotations were designed  to best match the
orientation of the characteristic six-pointed  diffraction
pattern as it appeared on the array for both object and PSF 
calibrator.  This pattern is fixed with respect to the telescope
and thus in paralactic angle and  therefore rotates in the field
as the telescope tracks.  The resulting PSF was thus not rotated
to equatorial north, but was rotated through a superposition of 
rotation angles corresponding to the orientation of the
diffraction  pattern of the object sequence as it appeared on
the array. The  rotations for each PSF/object combination were
always matched pair-wise,  producing a matched PSF for every
subset of object images  under consideration.

\subsection{FWHM Analysis}
Because the combined images of the objects were all very nearly
point-like  and closely resembled the combined PSF calibrator
images, and variations in the PSF, not statistical noise,
dominated the overall  uncertainty in deconvolving these images,
it was necessary to evaluate the stability of the PSF by 
comparing the raw object and PSF calibrator images before 
proceeding to image deconvolution. An estimate of the intrinsic
source size of the object was obtained by assuming the sizes
add in quadrature,

\hspace{2.0in}${\theta_{INT}}^2~=~{\theta_{OBJ}}^2~-~{\theta_{PSF}}^2$,
 
\noindent
where $\theta_{INT}$ is the intrinsic source size of the
object,  $\theta_{PSF}$ is the observed PSF size, and
$\theta_{OBJ}$ is  the observed object size all expressed as
FWHM.  An estimate of the uncertainty can be gotten from
${\delta\theta_{INT}}^2$, the 
dispersion of ${\theta_{INT}}^2$ in the population of pairs.
We will take the quantity
$\theta_{UL}~=~[<{\theta_{INT}}^2>~+~{\delta\theta_{INT}}^2]^{1/2}$ as
the estimate of the upper limit on the source size. It should be noted
that this definition takes account of the fact that the objects and PSFs 
are often comparable in size, and the fact that the size of the object 
image on occasion exceeds that of the PSF.

\subsection{Deconvolution}
Each combined object image was deconvolved by a matched PSF 
calibrator image using the Richardson-Lucy procedure (Richardson
1972; Lucy 1974) as encoded in the IDL maximum likelihood
routine (Varosi \& Landsman 1993).  The images were interpolated
to a four  times finer pixel scale before deconvolution.  The
reconvolved  image was rebinned and subtracted from the original
image and  compared with the estimated statistical uncertainty
per pixel.   It was found, however, that the uncertainties were
largest where the  image was brightest, and were always greater
than the statistical uncertainties even after iterating the
deconvolution routine a large number of  times. The non-uniform
uncertainties from deconvolution must be compared to a
combination of statistical and PSF uncertainties in the region
of image of interest.   Therefore we chose to stop iterating
the  deconvolution routine when the residuals were consistent
with the measurement uncertainties in the raw images.

\section{Results}
NGC~7469 showed a statistically significant  detection of
structure while the images of NGC~1275 and NGC~4151, combining
all the data, both showed a single point-like compact nucleus .
The FWHM of various angular quantities, obtained before
deconvolution was applied, are given in Table~3. Although the
uncertainties on the $<{\theta}_{INT}^2>^{1/2}$ are small, we
claim, to be conservative, only to have measured a limiting size
for the nuclei of NGC~1275 and NGC~4151 of $\theta_{UL}$.  It 
should be noted that  the
PSF during  the NGC~7469 observations was noticeably more
compact and  significantly more stable than for any of the other
objects. Deconvolution applied to  the combined NGC~7469 image confirmed
the presence of clear structure. 

In the  case of NGC~1275 and NGC~7469, the data were extensive
enough to provide  several sets of the object and the PSF 
calibrator.   The  images were paired to both match the object
and PSF calibrator  images in airmass and to minimize the time
interval between the two  images.  Unfortunately, the NGC~4151
data were quite limited, resulting in only two image pairs.  In
order to obtain a meaningful limit, the data were simply
sub-divided into five image pairs.  This is less than ideal 
since the pairs all have a common time interval, and the
matching in  airmass was poor.  Further sub-dividing the
NGC~4151 images was  possible, but gave a nearly  identical
result. From the present data we conclude that 0$''$.16 is an 
upper limit to the size of the nucleus. In comparison,
Neugebauer et al.\ (1990), using slit scans and a single element
detector, determined the size of the nucleus of NGC~4151 at
11.2~$\mu$m to be 0$''$.16~$\pm$~0$''$.04. The  deconvolved 
images of NGC~4151 and NGC~1275 do not show structure and  these
sources  appear to be unresolved in these observations. As
stated above, the deconvolution algorithm did not improve the
size limit beyond that set by the variations in the PSF and we
choose to quote the upper limit of the sizes as the size
estimated in  the analysis of the FWHM in the raw images. 

As expected from the work of Miles et al.\ (1996), the raw
images of NGC~7469 show a bright compact nucleus  surrounded by
a fainter ring. The flux from the  ring, which  has a
diameter of  $\sim$3$''$ and is well separated from the nucleus,
is comparable to that of the nucleus.  The surface brightness 
from the ring is less than 5\% of the brightness of the central
source  in the raw image. [Note that because of the small chop
amplitude of these observations we cannot make a reliable flux
measurement in an adequately large beam to determine this
ratio.] 

Photometric results are included in the summary of Table~4. With
20~\% uncertainties, the flux densities of the compact sources
of NGC~1275 and NGC~4151 reported here at 12.5~$\mu$m agree with
those measured at 12~$\mu$m by IRAS in a 4$'$.5~$\times$~0$'$.8
beam. These results are consistent with the results on NGC~4151
reported by Radomski et al.\ (2003), who find that 73~\% of the
10~$\mu$m broad band flux is in an unresolved component.  We
will assume that the luminosity and color as measured by IRAS
applies to the compact sources. The flux density of the compact
nucleus of NGC~7469 inside a 1\farcs4 diameter beam is 40~\% of
the  flux density at 12.0~$\mu$m measured in the much larger
IRAS beam. 

\subsection{Deconvolution of NGC~7469}  
  
The deconvolved image of NGC~7469, after combining all the data,
shows significant structure both in the ring and in the nucleus.
The deconvolution routine reaches statistical  noise on the
3$''$ diameter ring in about ten iterations, due to its low
surface  brightness.  Maximum likelihood does not perform well
on images  like this requiring large dynamic range, and we note
that although  many more iterations are needed to resolve the
nucleus, they are of  no further use in resolving the ring. The
deconvolution of the image with ten iterations, thus emphasizing
the 3$''$ diameter ring, is shown in Figure~1. For purposes of
deconvolving  the nucleus, the ring does not affect the images
of the  central source.  This assertion was  tested by
deconvolving a 3\farcs8~$\times$~3\farcs8 image of NGC~7469
containing the ring, and by deconvolving a
1\farcs6~$\times$~1\farcs6 image of NGC~7469 excluding the ring.
Both deconvolutions produce essentially identical images of the
inner 1$''~\times~1''$ nucleus. In the following we will often
designate the central nucleus of NGC~7469 simply  as
``NGC~7469''. 

The deconvolved image of the NGC~7469 nucleus with 1000
iterations of the deconvolution algorithm is presented in
Figure~2 and shows extended  structure that is statistically
significant based on the estimate of PSF  variations.  The
robustness of the deconvolution technique was tested by
sub-dividing the observations of the PSF into subgroups and
treating successive subgroups as objects. This test is described in
Appendix~A. The reproducibility of the   extension shown in
Figure~2 was likewise tested by sub-dividing the data into
independent  object and PSF calibrator images and deconvolving
the  corresponding image pairs, as shown in the Appendix~B. 
The statistical
significance can be evaluated quantitatively by comparing
the quality of fit of the hypothesized elongated source to that
obtained under the  assumption of a simple point source. This
procedure is described in Appendix~C with the result that the
elongation is  estimated to be significant at the 4.3-$\sigma$
confidence level. 

To summarize the appendices, we conclude the  extension of
($<$0\farcs4)~$\times$~0\farcs08 seen in NGC~7469 is real
for the following reasons: 

1. Structure in NGC~7469 was detected statistically based on the FWHM 
of the object and the PSF.

2. Deconvolved PSF images, synthesized from the observations for
matching in time and  airmass, are point-like and more compact
than the deconvolved  images of NGC~7469.

3. Deconvolved images using independent pairs of NGC~7469 and
the PSF  produce consistent results with a northwest-southeast
extension of similar size  and direction. 

\subsection{Extended Emission}

Although the observations were not optimized to study extended
emission, the quantity and quality of the data on NGC~1275 and
NGC~7469 were sufficiently great to attempt an examination of
the extended emission  using the technique exploited
successfully by Radomski  et al. \ (2003) in the study of
NGC~4151. These high resolution images, combined with a precise
knowledge of the PSF, allow us to best probe extended emission
close to the nucleus.  

In Figure~3, two sets of differences -- the galaxy minus the PSF
and the galaxy minus a fraction of the PSF --are given.  There
is evidence in NGC~7469 for emission in the northwest--southeast
direction --- i.e., the orientation of the structure in the
deconvolved image --- extending almost 1$''$ while in NGC~1275
there is an indication of extended emission  in the north-south
direction.  The putative extended emission is only about three
times brighter than the PSF residuals, so we advise caution in
interpreting these images, especially at the radius of the first
Airy maximum (radius = 0.4$''$).  Although faint, the extended
emission reproduces itself if the images are constructed using
independent subsets of the data from the first half and second
half of the observations, in spite of the fact that the change
in paralactic angle rotates the apparent diffraction pattern by
10 $\degr$. In NGC~7469, the structure noted here is also seen in
Figure~1.

\section{Discussion}

The observations at 12.5~$\mu$m with the 10-m Keck  Telescope
yield a resolved structure in NGC~7469, but give only limits for
the size of NGC~1275 and NGC~4151; see Table~4.  At the distances of the
galaxies, the limits on the angular sizes correspond to linear
size limits of 10~pc in NGC~4151 and 27~pc in NGC~1275. Although
an asymmetry is obvious in NGC~1068 at a comparable physical
scale, none is obvious in NGC~4151 in our images. The physical temperatures
implied for dust grains heated by a central source of
luminosity  are between 300~K and 450~K depending on the sizes
of the individual grains and their  composition. This is
substantially hotter than the observed color temperature, based
on IRAS 12~$\mu$m and 25~$\mu$m data, of 192~K and 201~K for
NGC~1275 and NGC~4151 respectively.  As shown in Table~4, the
limits on the sizes in these sources are not small enough to
require optically thick emission, but the lower bound on
$\tau_{12.5{\mu}m}$ is 0.25 in NGC~1275 and 0.06 in
NGC~4151 if the emission is thermal emission at the
color temperature determined from the IRAS 12~$\mu$m to
25~$\mu$m flux density ratio. This calculation, of course, requires
the reasonable assumption that the source structure is essentially
the same at 25 and 12~$\mu$m.

In the case of NGC~7469, the resolved structure provides strong
direct evidence for the dust thermal emission model in the
nucleus as well as a limit of $\tau_{12.5{\mu}m}>$~0.4 within
this structure (Table~4). The elongated structure is suggestive
of a disk geometry. The ``standard model''of an active galactic
nucleus (AGN) (e.g., Krolik 1999; Sanders et al. 1989) suggests
that this disk is heated by the central AGN. The fact, noted
above, that graphite and silicate dust grains at distances of
$\sim$15~pc from the central source, i.e. at the linear radius
corresponding to the angular size of the structure, are heated
to temperatures of 300 - 450~K, well above the observed color
temperature, is consistent with this picture. The derived
brightness  temperature, again well below 300~K, is only a limit
since the  source is unresolved in one dimension.

An interesting question is what is the relationship of the
infrared nucleus to the nuclear structure of the galaxy observed
in the radio. Observations by Thean et al.\ (2001) at 1.6-GHz
(18-cm) at $\sim$~0\farcs3 resolution show structure whose
orientation is very similar to that seen in the mid-infrared
structure. A similar feature is also seen at 4.9-GHz (6-cm) in
the VLA image of Wilson et al.\ (1991). Although again at a much
lower spatial resolution (2$''$) than that seen here in the
mid-infrared, high resolution maps in CO (Meixner et al.\ 1990)
show emission, interpreted as a rotating molecular disk, whose
orientation also is similar to that seen in the mid-infrared
structure. 

VLBI observations of Lonsdale, Lonsdale \& Smith (2003) at 18-cm
show that the nucleus of NGC~7469 is a nearly symmetric linear
triple source aligned in the east-west direction with the outer
components separated by 0\farcs15, and the central source nearly
centered between these two components.  The agreement between
the centroids of the radio peak, as described by Lonsdale et
al.\ , and the mid-infrared peaks is excellent, the two peaks
coinciding to within 0\farcs040 if the 3$''$ ring structures are
assumed to be spatially coincident. The uncertainty in the
determination of the astrometric coincidence of the radio and
infrared nuclei is determined by how well the structures in the
3$''$ ring can be aligned.

 We have compared the mid-infrared image of Figure~2 with the VLBI image of Lonsdale et al.(2003) in Figure 4. The position angle of the elongated
structure at 12.5~$\mu$m is 135\degr, $\sim$45\degr ~off of the
position angle of the linear radio structure. To
register the images in Figure~4 we assumed that the middle
radio peak coincides with the AGN nucleus, and is centered on
the mid-infrared peak. Then the two outer radio components are
symmetrically placed with respect to the mid-infrared disk, and
is suggestive of radio jets emerging out of the plane of the
dust disk. If the galaxy is assumed to be circular, its apparent
size on the sky -- 1.1' $\times$ 1.5' (de~Vaucouleurs et al. 1991)
-- implies it is at an inclination angle of $\sim$45\degr. 

The surface luminosity densities  of the three compact nuclei,
as listed in Table~4, are on the order of a few times  
10$^{14}$~L$_\sun$~Kpc$^{-2}$. Soifer et al.\ (2000) found that,
of seven Ultraluminous Infrared Galaxies (ULIRGs) observed in
the mid-infrared with the Keck 10-m Telescope, only
Markarian~231, a known Seyfert~1 galaxy, has a luminosity
density as high as 10$^{14}$~L$_\sun$~Kpc$^{-2}$. The difference
between starbursts and AGNs is heightened by an examination of
the surface brightnesses of infrared luminous  galactic nuclei
listed by Soifer et al.\ , Soifer et al.\ (2001), and Evans et
al.\ (2003). Galactic HII regions have surface brightnesses
which range from 2~$\times$~10$^{11}$ to 2~$\times$~10$^{12}$
L$_\sun$Kpc$^{-2}$, those of the starburst galaxies studied
range from 2~$\times$~10$^{11}$ to 10$^{13}$ L$_\sun$Kpc$^{-2}$,
while those of the ULIRGS, with the exception of Markarian~231,
range from 2~$\times$~10$^{12}$ to 6~$\times$~10$^{13}$
L$_\sun$Kpc$^{-2}$. Gorjian, Turner \& Beck (2001) have reported
the infrared surface brightness of a ``super'' star cluster of
$\sim$~3~$\times$~10$^{14}$~L$_\sun$~Kpc$^{-2}$ over  1~--~2~pc
in the nucleus of NGC~5253. While the peak surface brightness in
the ``super'' star cluster matches that of the Seyfert galaxies
the physical extent of the ``super'' star cluster is smaller, 
consistent with the 1~--~2 orders of magnitude lower
luminosities compared to the Seyfert nuclei. Thus it is seen
that the infrared surface brightness of those galaxies with
evidence of AGN activity is significantly higher than that of
those dominated by star formation and we can conclude that star
formation provides only a negligible fraction on the luminosity
of the nuclei of the Seyfert galaxies measured here. These
results argue that the infrared surface brightness, in concert
with the infrared luminosity, can be used as a diagnostic
measure for the presence of starbursts or AGNs. These
observations can be used to measure the relative contributions
of AGN and starbursts to the energy budgets of infrared luminous
galaxies provided that the AGN is not heavily obscured at the
observed wavelength.

\section{Summary}

High spatial resolution observations  at 12.5~$\mu$m were made
with the 10-meter diameter Keck Telescope  in order to study the
central nuclei of three Seyfert galaxies.

1) The observations, when deconvolved,  resolve a linear
structure in the Seyfert galaxy NGC~7469 that is
($<$13)~$\times$~26~pc. The orientation of the infrared emission, when
compared to CO and radio observations, supports the conjecture
that the mid~-~infrared emission traces a disk. 

2) The observations do not detect structure in the nuclei of
NGC~1275 (FWHM~$<$~28~pc) or of NGC~4151 (FWHM~$<$~10~pc).  The
former limits are within a factor of two of being able to
resolve  a significant contribution to the total emission by 
thermal emission of centrally heated dust grains.

3) There are indications of underlying extended emission of
about 300~pc in NGC~7469 and NGC~1275. 

4) The luminosity surface brightness derived for the three
Seyfert galaxies are a few times 10$^{14}$ L$_\sun$Kpc$^{-2}$,
significantly higher than that associated with starbursts.

\acknowledgments{}  
We thank the staff of the Keck Observatory for their assistance
in making these observations possible.  The W.  M. Keck
Observatory is operated as a scientific partnership between the
California Institute of Technology, the University of California
and the National Aeronautics and Space Administration. It was
made possible by the generous financial support of the W. M.
Keck Foundation.  We extend special thanks  to those of Hawaiian
ancestry on whose sacred mountain we are privileged to be
guests.  Without their generous hospitality, none of the
observations presented herein would have been possible. We thank
Carol Lonsdale for supplying the VLBI.image of NGC~7469. T.S.
and E.E. were supported by grants from the NSF and  NASA.
B.T.S. and L.A. are supported by the SIRTF Science Center at
Caltech.  SIRTF is carried out at J.P.L., operated by Caltech
under an agreement with NASA.

\newpage
\begin{center}
\textbf{\hspace{1.3in}APPENDICES}
\newline
\textbf{A - PSF STABILITY}
\end{center}

The stability of the PSF accompanying NGC~7469 was estimated by
deconvolving a synthesized  image of the PSF calibrator by a
complementary set of independent  images of the PSF calibrator. 
It is important to accurately match the  two PSF images in time
and airmass.  When deconvolving the NGC~7469  image, the average
airmass of the object differs only by 0.005  from the average
airmass of the PSF, and the average time interval  between
object and PSF sub-images is eight minutes. 

Subgroups to serve as the object and PSF were synthesized using
the fact that each of the four sub-images of the PSF  actually
consists of four independent chop-nod sets; in Table~A1 these
are called a, b,  c, and d. The composition of several
synthesized PSFs is shown in  Table A1. Each pair of PSFs are
independent, and were chosen with variable  matching of the PSF
pair in time and airmass.   Deconvolved images resulting from
the use of one subgroup as PSF and one as object  are
illustrated in Figure~5.  Close  matching in time and airmass
(e.g., PSF\_C//PSF\_D) produce point-like deconvolved images.
Poor matching in airmass (e.g., PSF\_12//PSF\_34) results in
significant extensions. Matching in airmass seems to be 
somewhat more important than matching in time, as the
PSF\_13//PSF\_24 and PSF\_12//PSF\_34 pairs produce the most
extended  deconvolved images.  For example, the pair
PSF\_14//PSF\_23 has  the same time interval as the pair
PSF\_12//PSF\_34 but much better matching  in airmass. The most
representative matching for purposes  of deconvolving the object
are PSF\_C//PSF\_D. 

\begin{center}
\textbf{B - NGC~7469}
\end{center}

The image of NGC~7469 was deconvolved by  combining all of the
object and PSF data.  The object and PSF images  are very
closely matched in airmass and time interval (see Table~A2).  
The data set was divided into independent images of the object 
and PSF to assess the stability of the northwest-southeast
extension in the  deconvolved image of NGC~7469.  Splitting the
data into two  independent images of NGC~7469 and the PSF,
N\_12//PSF\_12 and  N\_34//PSF\_34, results in two images which
both show a similar northwest-southeast   extension; see
Figure~6.  This agreement is remarkable given that the image
pairs  are completely independent. As shown in Figure~6, further
sub-dividing the data into four sub-images and PSFs results in
four individual images which all show  the northwest-southeast
extension.  Furthermore, all images also show two smaller 
isolated structures located about 0$''$.5 to the northwest and
southwest.  A smaller  structure located $\sim$0$''$.3 to the
east is seen in the two image pairs, but  not in each of the
four independent sub-images.

\begin{center}
\textbf{C - Statistical Significance of the Structure}
\end{center}

The statistical significance of the elongated structure apparent
in the deconvolved images of NGC~7469 was evaluated by comparing
the quality of fit of the hypothesized elongated source to that
obtained under the  assumption of a simple point source. This
comparison is based on the residuals with respect to the
observed data for the two cases, taking into account our
knowledge of the properties of the  measurement noise.  The
residual images are shown in panels (e) and (f) of Figure~7,
along with the sequence of images used to generate them,
consisting of: (a) observed  image, (b) PSF, (c) ``model" image
representing the central source obtained by deconvolution, and
(d) the model image convolved with the PSF. The residual image
(e) was  obtained by subtracting (d) from (a), and is to be
compared with (f), which represents the residual image
corresponding to the best  point-source fit, i.e, the residual
image obtained by subtracting (b) from (a). The comparison of
(e) and (f) shows that the residuals are noticeably lower for
the extended source than for the point source. 

The relative probabilities of the two source-geometry
hypotheses  can be calculated by comparing the  conditional
probabilities of the extended source and a point source given
the data (represented by a vector {\bf y} whose components
represent the pixel values of the observed image) by $P({\rm
ext}|{\bf y})$ and $P({\rm point}|{\bf y})$. A simple
application of Bayes' rule then yields:
\begin{equation}
{\rm ln}\,\,P({\rm ext}|{\bf y})/P({\rm point}|{\bf y}) =
\frac{1}{2}\sum_i [(y_i-c^{\rm point}_i)^2 - (y_i-c^{\rm ext}_i)^2]/
\sigma_i^2
\end{equation}
where $c^{\rm point}_i$ and $c^{\rm ext}$ represent the
point-source and extended-source model images convolved with the
PSF, respectively, and $\sigma_i$ represents the measurement
noise in the $i$th pixel.  The summation is over all
statistically independent pixels in the image.  In deriving this
equation a flat prior was assumed, representing equal  {\em a
priori\/} probabilities of extended and point sources.

The measurement noise is assumed to be a Gaussian random process with
variance:
\begin{equation}
\sigma_i^2 = \sigma_{\rm bg}^2 + f^2 V_i
\end{equation}
where $\sigma_{bg}$ is the standard deviation of the background
(obtained using pixels far from the source itself), $f$ is the
source strength in data units, and $V_i$ is the
``variance map" of the PSF,  representing the normalized value
of PSF uncertainty as a function of position. The PSF variance map was
obtained using the deconvolution residuals of four independent
sub-images of the PSF calibrator and smoothing the result using
a $5\times 5$-pixel boxcar averaging window. The PSF uncertainty
term in the above expression represents a good approximation in
the case of a marginally resolved source.

Based on this procedure, a probability ratio $P({\rm point}|{\bf y})/P({\rm
ext}|{\bf y}) = 1.5 \times 10^{-5}$ was obtained.  After normalizing the sum
of the two probabilities to unity, this result implies that the probability
that the data were produced by a point source was of this same order, which
corresponds to a $4.3 \sigma$ (or greater) deviation of a Gaussian random
variable.  We can therefore equate the significance of our result to a
$4.3\sigma$ detection of source elongation.
\newpage
\figurenum{1}
\figcaption{ 
The deconvolved image of NGC~7469 is shown. There were only ten
iterations in the deconvolution, so the faint ring with
$\sim$3$''$ diameter is emphasized. The image is  displayed with
contours spaced by multiplicative factors of 1.342, ranging 
from 90\% of maximum so that the third contour represents the
50\% level. North is up and east is to the left. 
}

\figurenum{2}
\figcaption{
The deconvolved image of NGC~7469 is shown with 1000 iterations
in the deconvolution thus emphasizing the nucleus. The image is 
displayed with contours spaced by multiplicative factors of
1.342, ranging  from 90\% of maximum so that the third contour
represents the 50\% level. North is up and east is to the left.
} 

\figurenum{3}
\figcaption{ 
The raw PSF is plotted in the top row for NGC~7469 and
NGC~1275.  The object image minus 0.9 (in the case of NGC~7469)
or 0.95 (in the case of NGC~1275) times the PSF is plotted in
the second row and the  object minus PSF is plotted in the third
from top row. The number of  contours is the same in each of the
top three rows. PSF residuals obtained by differencing half the
PSF images against the other half are plotted in the bottom row.
North is up and east to the left. All images are smoothed with a
0\farcs3 FWHM  Gaussian beam to emphasize the extended emission,
and plotted with contours separated by multiplicative intervals
of $\sqrt{2}$.  The PSF is normalized to the peak of the object
image before subtraction.   The PSF residuals are scaled such
that the bottom contour, as a fraction of PSF flux, matches the
bottom contour plotted in the object minus PSF panel, as a
fraction of object flux.  The two PSFs images used to generate
the PSF residuals image were carefully selected to mimic the
matching between the PSF and object.  Since the amount of data
used to generate the each PSF is halved, we expect that the
residuals are overestimated.  
}

\figurenum{4}
\figcaption{
The contours of the deconvolved image of NGC~7469 at 12.5~$\mu$m, 
as given in Figure~2, are overlaid on the VLBI image of
Lonsdale et al.\ (2003). The central source of the VLBI image has
been assumed to coincide with the infrared source.
}
\figurenum{5}
\figcaption{ 
Deconvolved images of the PSF accompanying NGC~7469, synthesized
as listed in Table~A1, are given; see Appendix~A.  The images
are all  displayed with contours spaced by multiplicative
factors of two, ranging  from 90\% of maximum to 0.35\% of
maximum. North is up and east is to the left.  
}
\figurenum{6}
\figcaption{  
The deconvolved image of NGC~7469 combining all of the data  is
given in the top image and is compared to independent data
subsets in the lower images; see Table~A2.  Using the first two 
images of the object and PSF, and the second two images of the 
object and PSF, result in N\_12//PSF\_12 and N34//PSF\_34 
respectively.  These images bear a strong resemblance to the 
N\_T//PSF\_T image from the total data set.  Further
sub-divisions of the  images into four independent subsets
(bottom four images) are also  shown.  The images are all
displayed with contours spaced by  multiplicative factors of
two, ranging from 90\% of maximum to 0.35\% of maximum.  North
is up and east is to the left. 
}

\figurenum{7}
\figcaption{  
(a) Observed (shift and add) image of NGC~7469; (b) PSF image; 
(c) Model image of the extended source (representing the central
portion of the deconvolved image); (d) Convolution of model image with PSF;
(e) Residual image, obtained by subtracting (d) from (a); (f) Residual image
obtained from the best fit to a point source, for comparison.  All images
have a field of view of 1\farcs28$\times$1\farcs28 and are presented on a 
sampling grid whose spacing is 0.25 focal-plane pixels.   The intensity scale
is logarithmic (2-decades) for (a)--(d) and linear for (e) and (f).
}

\clearpage

\normalsize
\singlespace

\newpage

\begin{deluxetable}{ccccccc}
\tablewidth{0pt}
\tablenum{1}
\tablecaption{Sample}
\tablehead{
\colhead{Object}               &\colhead{Seyfert}      &
\colhead{Redshift}             &\colhead{Scale}        &
\multicolumn{2}{c}{IRAS~f$_\nu$}                       &
\colhead{log[L$_{IR}$(L$_\sun$)]\tablenotemark{a}}     \\
\colhead{}                     &\colhead{Type}         & 
\colhead{}                     &\colhead{}             &          
\colhead{12~$\mu$m}            &\colhead{25~$\mu$m}    &    
\colhead{}                                             \\
\colhead{}                     &\colhead{ }            &
\colhead{}                     &\colhead{pc/$''$}      &
\colhead{Jy}                   &\colhead{Jy}           &          
\colhead{}                                             
}
\startdata
NGC~1275 & 2   & 0.0176\tablenotemark{b}&350 & 1.17 & 3.50 & 11.24 \\
NGC~4151 & 1.5 & 0.0033\tablenotemark{c}& 65 & 1.95 & 5.04 &  9.86 \\
NGC~7469 & 1.2 & 0.0163\tablenotemark{d}&326 & 1.60 & 5.40 &
11.59\tablenotemark{e} \\
\enddata
\tablenotetext{a}{Based on IRAS observations from 12 to
100~$\mu$m}
\tablenotetext{b}{Strauss et al.\ (1992)}
\tablenotetext{c}{de Vaucouleurs et al.\ (1991)}
\tablenotetext{d}{Keel (1996)}
\tablenotetext{e}{Luminosity of both nucleus and 3$''$ diameter
ring}

\end{deluxetable}

\newpage

\begin{deluxetable}{cccccc}
\tablewidth{0pt}
\tablenum{2}
\tablecaption{Log of Observations}
\tablehead{
\colhead{Object}                            &
\colhead{Date}                              &
\colhead{PSF}                               &
\multicolumn{2}{c}{Integration}             &
\colhead{Number}                            \\
\colhead{}                                  &
\colhead{}                                  &
\colhead{Star}                              &
\multicolumn{2}{c}{Time\tablenotemark{a}}   &
\colhead{of PSF}                            \\           
\colhead{}                                  &
\colhead{}                                  &
\colhead{}                                  &
\colhead{Object}                            &
\colhead{PSF}                               &
\colhead{Comparisons}                       \\
\colhead{}                                  &
\colhead{2000}                              & 
\colhead{}                                  &
\multicolumn{2}{c}{s}                       &
\colhead{}
}
\startdata
NGC~1275 & Dec & IRC +40060  & 30 & 27 & 9 \\
NGC~4151 & May & HR~4550     & 54 & 54 & 2 \\
NGC~7469 & Sep & HR~8815     & 81 & 54 & 4 \\ 
\enddata
\tablenotetext{a}{Integration time in each comoarison}

\end{deluxetable}

\newpage
\begin{deluxetable}{ccccc}
\tablewidth{0pt}
\tablenum{3}
\tablecaption{FWHM Analysis}
\tablehead{
\colhead{Object}                                        &
\colhead{$<\theta_{OBJ}>$\tablenotemark{a}}             &
\colhead{$<\theta_{PSF}>$\tablenotemark{a}}             & 
\colhead{$<{\theta_{INT}^2}>^{1/2,~}$\tablenotemark{b}} &
\colhead{$\theta_{UL}$ }                                \\ 
\colhead{}                                              & 
\multicolumn{4}{c}{$''$}                                                                         
}
\startdata
NGC~1275 & 0.35 $\pm$0.02  & 0.34 $\pm$0.02  & 0.06~+0.02~-0.03 &  0.08\\
NGC~4151 & 0.38 $\pm$0.02  & 0.36 $\pm$0.05  & 0.10~+0.06~-0.10 &  0.16\\
NGC~7469 & 0.276$\pm$0.006 & 0.270$\pm$0.003 & 0.070$\pm$0.005 &  0.08\\
\enddata

\singlespace
\tablenotetext{a}{uncertainties are sample standard
deviation of $\theta_{OBJ}$ or $\theta_{PSF}$}
\tablenotetext{b}{uncertainties calculated from standard deviation 
in $<{\theta_{INT}^2}>$}

\end{deluxetable}

\newpage

\begin{deluxetable}{ccccccccc}
\rotate
\tablewidth{0pt}
\tablenum{4}
\tablecaption{Results and Derived Quantities}
\tablehead{
\colhead{Object}                             &
\multicolumn{2}{c}{f$_\nu$}                  & 
\multicolumn{2}{c}{Size\tablenotemark{a}}    &
\colhead{T$_{Brightness}$}                   &
\colhead{T$_{Color}$}                        &
\colhead{Diameter}                           &
\colhead{Surface}                            \\
\colhead{}                                   &
\colhead{12~$\mu$m\tablenotemark{b}}                          &
\colhead{12.5~$\mu$m}                        & 
\colhead{12.5~$\mu$m}                        &
\colhead{12.5~$\mu$m}                        &  
\colhead{12.5~$\mu$m}                        &         
\colhead{12/25}                              &
\colhead{T$_C$=T$_B$}                        &
\colhead{Brightness\tablenotemark{e}}        \\
\colhead{}                                   &
\multicolumn{2}{c}{Jy}                       &
\colhead{$''$}                               &
\colhead{pc}                                 &
\colhead{\degr}                              &         
\colhead{\degr}                              &
\colhead{$''$}                               &
\colhead{10$^{14}$~L$_\sun$~Kpc$^{-2}$}
}
\startdata
NGC~1275 & 1.17 & 1.00\tablenotemark{c}   & $<$0.08 & $<$27 & $>$150 &  
192 & 0.04 & $>$2.8 \\
NGC~4151 & 1.95 & 2.31\tablenotemark{c}   & $<$0.16 & $<$10 & $>$139 &  
201 & 0.04 & $>$ 0.9 \\
NGC~7469 & 1.60 & 0.65\tablenotemark{d}   & ($<$0.04)$\times$0.08 &
($<$13)$\times$26 & $>$148 & 185 & 0.03 & $>$ 4.6\tablenotemark{f} \\
\enddata
\singlespace
\tablenotetext{a}{ limits are $\theta_{UL}$ }
\tablenotetext{b}{ For the convenience of the reader, the flux
densities obtained by IRAS at 12~$\mu$m are repeated here}
\tablenotetext{c}{ 4$''$ diameter beam; raw image; not deconvolved}
\tablenotetext{d}{ 1.4$''$ diameter beam; raw image;  not deconvolved}
\tablenotetext{e}{ 12.5~$\mu$m surface area; infrared luminosity}
\tablenotetext{f}{ big beam (IRAS) luminosity value $\times$~0.4}

\end{deluxetable}

\newpage

\begin{deluxetable}{ccccccccccccccccccc}
\tablewidth{0pt}
\tablenum{A1}
\tablecaption{Synthesized PSF Images}
\tablehead{
\colhead{PSF}              &
\multicolumn{4}{c}{1}      &
\multicolumn{4}{c}{2}      &
\multicolumn{4}{c}{3}      &
\multicolumn{4}{c}{4}      &
\colhead{$\Delta$AM}       &
\colhead{$\Delta$t}        \\
\colhead{}                 &
\colhead{a}                &
\colhead{b}                &
\colhead{c}                &
\colhead{d}                &
\colhead{a}                &
\colhead{b}                &
\colhead{c}                &
\colhead{d}                &
\colhead{a}                &
\colhead{b}                &
\colhead{c}                &
\colhead{d}                &
\colhead{a}                &
\colhead{b}                &
\colhead{c}                &
\colhead{d}                &
\colhead{}                 &
\colhead{min}              \\
}
\startdata
 A&$\surd$&$\surd$&&&$\surd$&$\surd$&&&$\surd$&$\surd$&&&$\surd$&$\surd$&&&0.000&3\\

B&&&$\surd$&$\surd$&&&$\surd$&$\surd$&&&$\surd$&$\surd$&&&$\surd$&$\surd$&&\\
\hline
 C&$\surd$&&&&$\surd$&$\surd$&$\surd$&&&$\surd$&$\surd$&$\surd$&&&&$\surd$&0.002&10\\
 D&&$\surd$&$\surd$&$\surd$&&&&$\surd$&$\surd$&&&&$\surd$&$\surd$&$\surd$&&&\\
\hline
13&$\surd$&$\surd$&$\surd$&$\surd$&&&&&$\surd$&$\surd$&$\surd$&$\surd$&&&&&0.025&18\\
24&&&&&$\surd$&$\surd$&$\surd$&$\surd$&&&&&$\surd$&$\surd$&$\surd$&$\surd$&&\\
\hline
14&$\surd$&$\surd$&$\surd$&$\surd$&&&&&&&&&$\surd$&$\surd$&$\surd$&$\surd$&0.005&30\\
23&&&&&$\surd$&$\surd$&$\surd$&$\surd$&$\surd$&$\surd$&$\surd$&$\surd$&&&&&&\\
\hline
12&$\surd$&$\surd$&$\surd$&$\surd$&$\surd$&$\surd$&$\surd$&$\surd$&&&&&&&&&0.045&30\\
34&&&&&&&&&$\surd$&$\surd$&$\surd$&$\surd$&$\surd$&$\surd$&$\surd$&$\surd$&&\\
\enddata

\end{deluxetable}

\newpage

\begin{deluxetable}{ccccccccccc}
\tablewidth{0pt}
\tablenum{A2}
\tablecaption{NGC~7469 Deconvolution Pairs}
\tablehead{
\colhead{Object\tablenotemark{a}} &
\multicolumn{2}{c}{1}      &
\multicolumn{2}{c}{2}      &
\multicolumn{2}{c}{3}      &
\multicolumn{2}{c}{4}      &
\colhead{$\Delta$AM}       &
\colhead{$\Delta$t}        \\
\colhead{}                 &
\colhead{N}                &
\colhead{P}                &
\colhead{N}                &
\colhead{P}                &
\colhead{N}                &
\colhead{P}                &
\colhead{N}                &
\colhead{P}                &
\colhead{}                 &
\colhead{min}              \\
}
\startdata
 N\_T      &  $\surd$& &  $\surd$& &  $\surd$& &  $\surd$& &  0.005& 8 \\
 PSF\_T    &   &$\surd$&   &$\surd$&   &$\surd$&   &$\surd$&       &   \\
 \hline
  N\_12    &  $\surd$& &  $\surd$& &   & &   & &  0.005& 8 \\
 PSF\_12   &   &$\surd$&   &$\surd$&   & &   & &       &   \\
 \hline
 N\_34     &   & &   & &  $\surd$& &  $\surd$& &  0.005& 8 \\
 PSF\_34   &   & &   & &   &$\surd$&   &$\surd$&       &   \\
 \hline
  N\_1     &  $\surd$& &   & &   & &   & &  0.000& 8 \\
 PSF\_1    &   &$\surd$&   & &   & &   & &       &   \\
 \hline
 N\_2      &   & &  $\surd$& &   & &   & &  0.010& 8 \\
 PSF\_2    &   & &   &$\surd$&   & &   & &       &   \\
 \hline
  N\_3     &   & &   & &  $\surd$& &   & &  0.000& 8 \\
 PSF\_3    &   & &   & &   &$\surd$&   & &       &   \\
 \hline
 N\_4      &   & &   & &   & &  $\surd$& &  0.010& 8 \\
 PSF\_4    &   & &   & &   & &   &$\surd$&       &   \\ 
\enddata
\tablenotetext{a}{N~$\leftrightarrow$~NGC~7469; PSF~$\leftrightarrow$~PSF star HR~8815}

\end{deluxetable}


\begin{references}{ 
Beichman, C. A., Neugebauer, G., Habing, H. J., Clegg, P. E. \& Chester, 
T. J. 1985, Infrared Astronomical Satellite (IRAS) Catalog and Atlases, 
Explanatory Supplement (Washington, DC, Government Printing Office)

de Vaucouleurs, G., de Vaucouleurs, A., Corwin, H. G., Jr., Buta, R. J., 
Paturel, G. \& Fouque, P. 1991, Third Reference Catalogue of Bright 
Galaxies Springer-Verlag Berlin Heidelberg New York)

Evans, A. S., et al. 2002, submitted,

Gorjian, V., Turner, J. L. \& Beck, S. C. 2001, ApJ, 554, L29

Joint IRAS Science Team 1989, IRAS Point Source
Catalog, Version 2 (Washington, DC, US Government Printing
Office)

Jones, B. \& Puetter, R. C. 1993, Proc. SPIE, 1946, 610

Keel, W. C. 1996, ApJS, 106, 27

Krolik, J. H. 1999, Active Galactic Nuclei: From the Central 
Black Hole to the Galactic Environment (Princeton, New Jersey, 
Princeton University Press)

Lonsdale, C. J., Lonsdale, C. J., Smith, H. E. \& Diamond, P. J.
2003, submitted  

Lucy, L. B. 1974, AJ, 79, 745

Meixner, M., Puchalsky, R., Blitz, L., Wright, M. \& Heckman, T. 
1990, ApJ, 354, 158

Miles, J. W., Houck, J. R., Hayward, T. L. \& Ashby, M. L. N. 
1996, ApJ, 465, 191

Neugebauer, G., Graham, J. R., Soifer, B. T. \& Matthews, K. 
1990, AJ, 99, 1456

Radomski, J. T., Pina, R. K., Packham, C., Telesco, C. M., 
Buizer, J. M. D., Fisher, R. S. \& Robinson, A. 2003, ApJ, 
(astro-ph/0212307) 

Richardson, W. H. 1972, JOSA, 62, 55

Sanders, D. B., Phinney, E. S., Neugebauer, G., Soifer, B. T. 
\& Matthews, K. 1989, ApJ, 347, 29

Seyfert, C. K. 1943, ApJ, 97, 28

Soifer, B. T., et al. 2000, AJ, 119, 509

Strauss, M. A., Huchra, J. P., Davis, M., Yahil, A., Fisher, K. B. 
\& Tonry, J. 1992, ApJS, 83, 29

Thean, A. H. C., Gillibrand, T. I., Pedlar, A. \& Kukula, M. J. 
2001, M.N.R.A.S., 327, 369

Varosi, F. \& Landsman, W. B. 1993, in Astronomical Data Analysis 
Software and Systems II, Vol. 52, ed. Hanisch, R. J., Brissenden, 
R. J. V. and Barnes, J. 

Wilson, A. S., Helfer, T. T., Haniff, C. A. \& Ward, M. J. 1991, 
ApJ, 381, 79

}

\end{references}
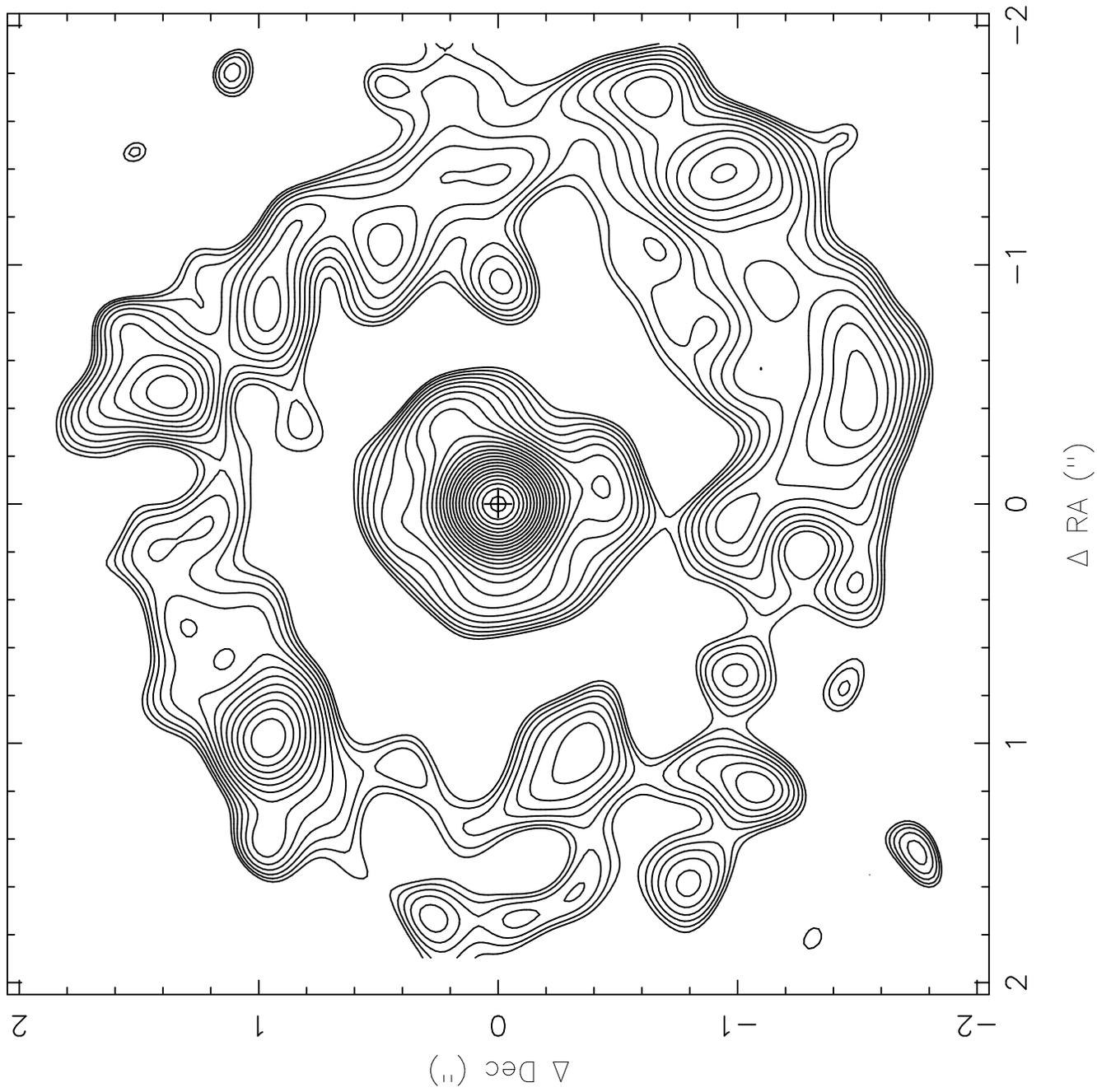
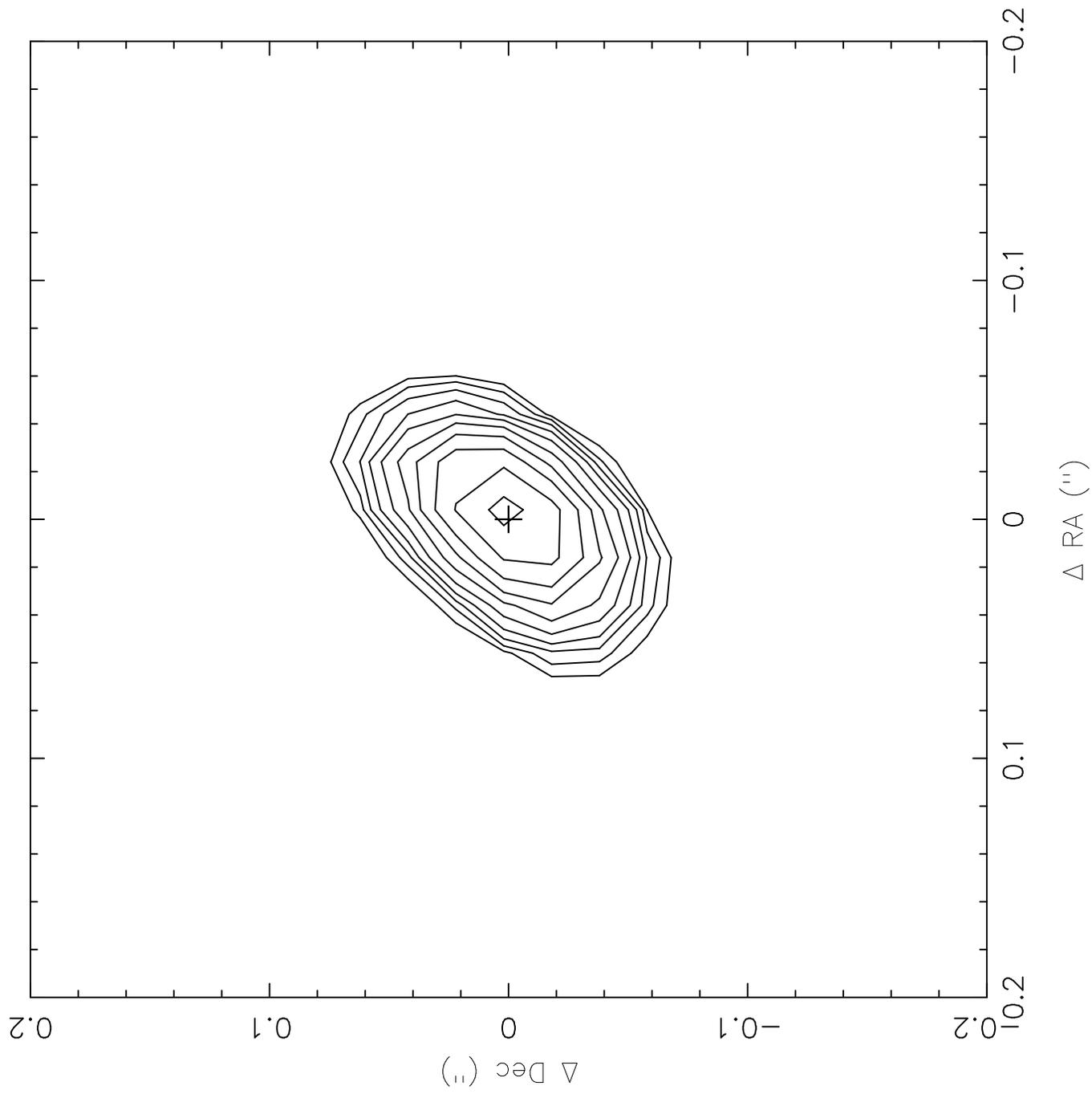
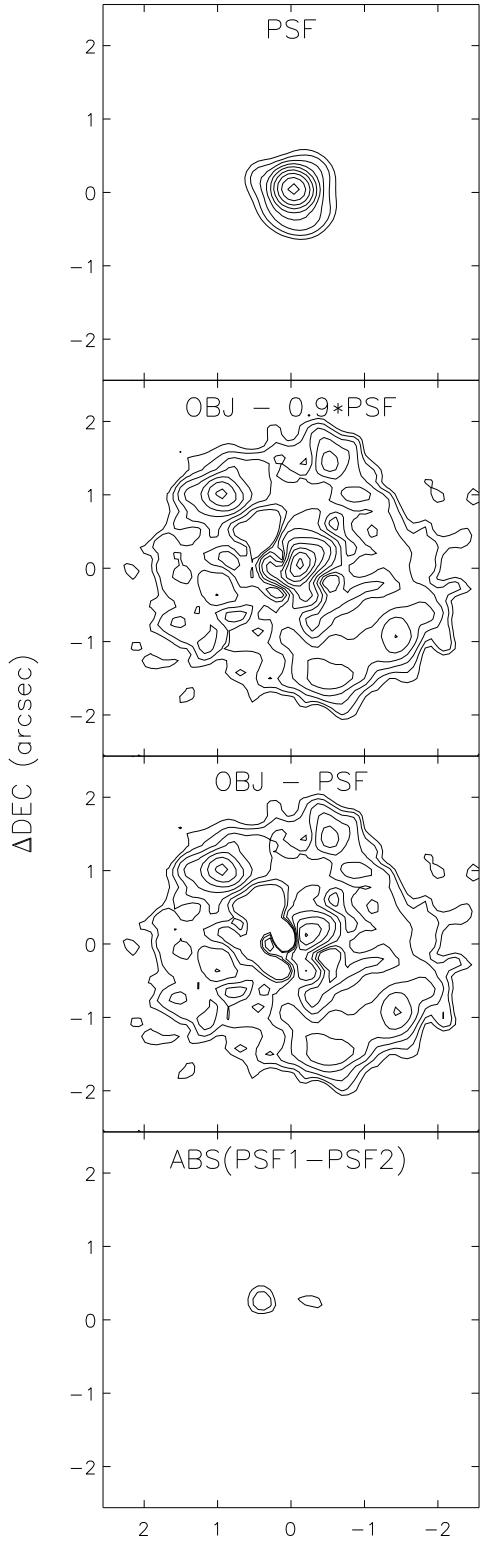
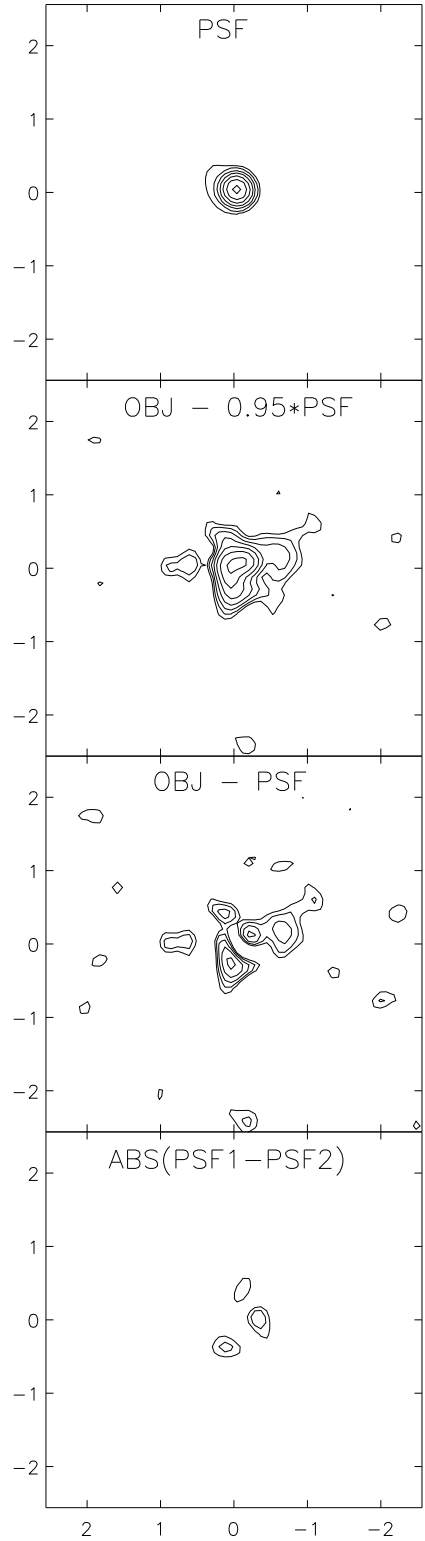
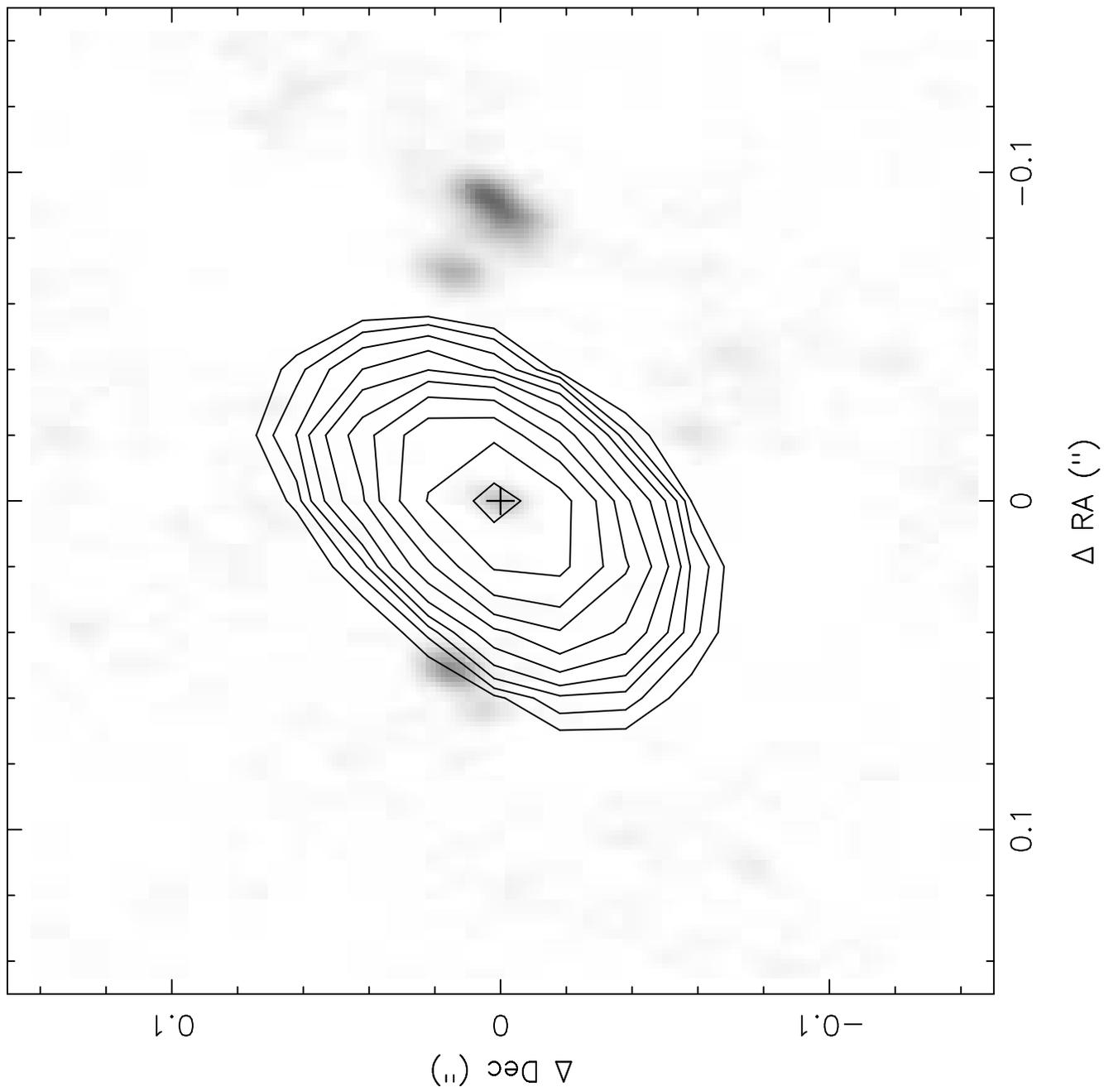
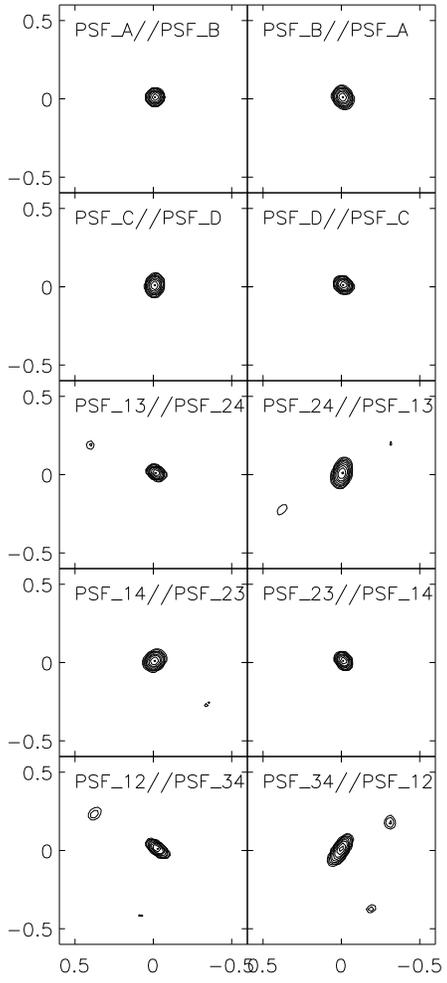
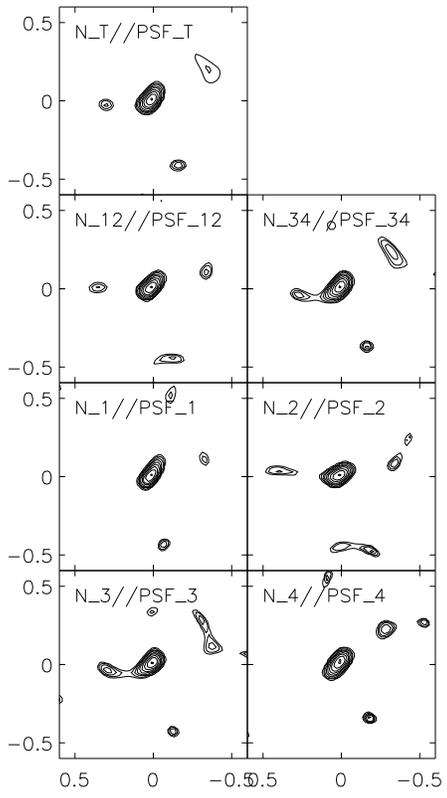
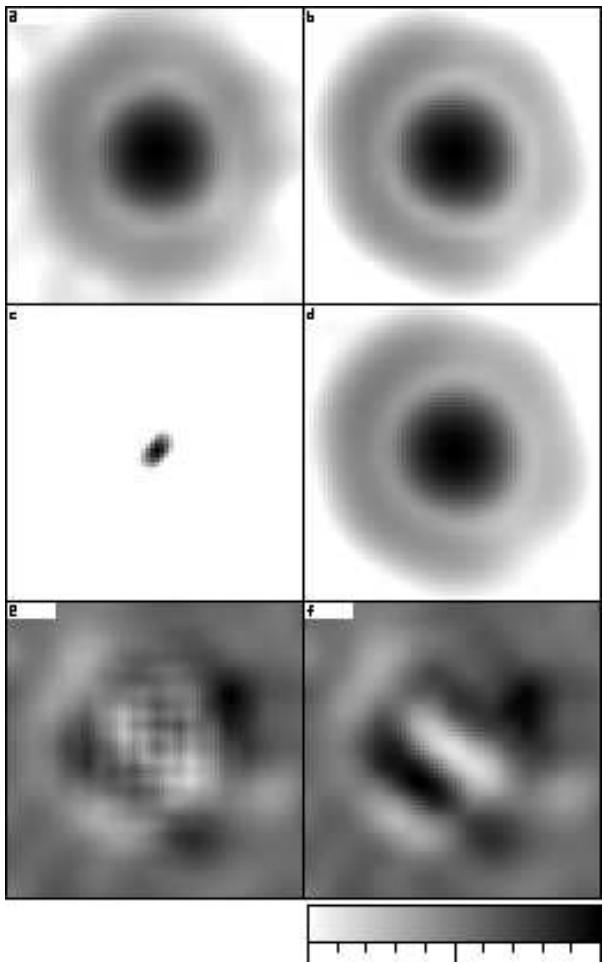
\end{document}